\documentclass[11pt]{article}

\usepackage[margin=1in]{geometry}
\usepackage{amsmath, amssymb, booktabs, graphicx, float, placeins}
\usepackage{caption}
\usepackage[numbers]{natbib}
\usepackage[hidelinks]{hyperref}

\graphicspath{{./}}

\title{Efficient Frontier Optimization of NBA Shot Selection Using Markov Reward Processes}
\author{Nirek Duma \and Leo Benaharon }
\date{}

\begin{document}
\maketitle

\begin{abstract}
This paper asks whether modern NBA shot selection can be evaluated as a dynamic portfolio problem rather than as a collection of isolated shot attempts.  We combine a possession-level Markov reward process with both a mean-variance shot allocation objective and a defense response cost.  The model separates rim attempts, midrange attempts, corner threes, and non-corner threes, then values each candidate shot diet by immediate scoring, continuation after offensive rebounds, empirical game-level variance, and a timeout-response stability cost.  Because fastbreak value depends on geometry that is not fully observed in standard play-by-play, we optimize the no-fastbreak sample first and then add back observed fastbreak shot selection as a fixed transition component. Our results indicate that the league aggregate operates near the efficient frontier, with marginal improvements to be made by slightly decreasing perimeter volume and mid range attempts while increasing rim attempts. Team-level analyses reflect this general trend. Ultimately, these findings suggest the three-point revolution has pushed volume slightly past its optimal boundary. By incorporating outcome variance and continuation effects, we show that optimizing solely for expected value mathematically overstates the benefit of perimeter shots.
\end{abstract}

\section{Introduction}

Over the past two decades, the NBA has greatly increased both pace and three-point attempt share.  The league did not only change where shots came from; it changed the rhythm of the game.  More possessions and more threes created the conditions for extreme scoring outbursts, including teams scoring over 150 points in regulation environments, but they also created a version of basketball that can become difficult to watch when variance turns the wrong way.  The clearest example is the 2018 Houston Rockets missing 27 straight threes in Game 7 of the Western Conference Finals.

In the modern NBA, we have seen offensive philosophies based on analytics use a common expected value argument to justify their shot selection. Executives like Daryl Morey have built teams with offensive philosophies centered almost entirely on the rim and the arc, popularizing the term 'Moreyball.' Mathematically, this seems sound, as rim attempts and threes yield higher expected returns than midrange attempts. The question is whether expected value alone is sufficient. A shot diet also carries continuation value, game-to-game variance, and defensive response pressure. These factors determine whether a high-value diet is actually stable.

\begin{figure}[H]
\centering
\includegraphics[width=0.9\textwidth]{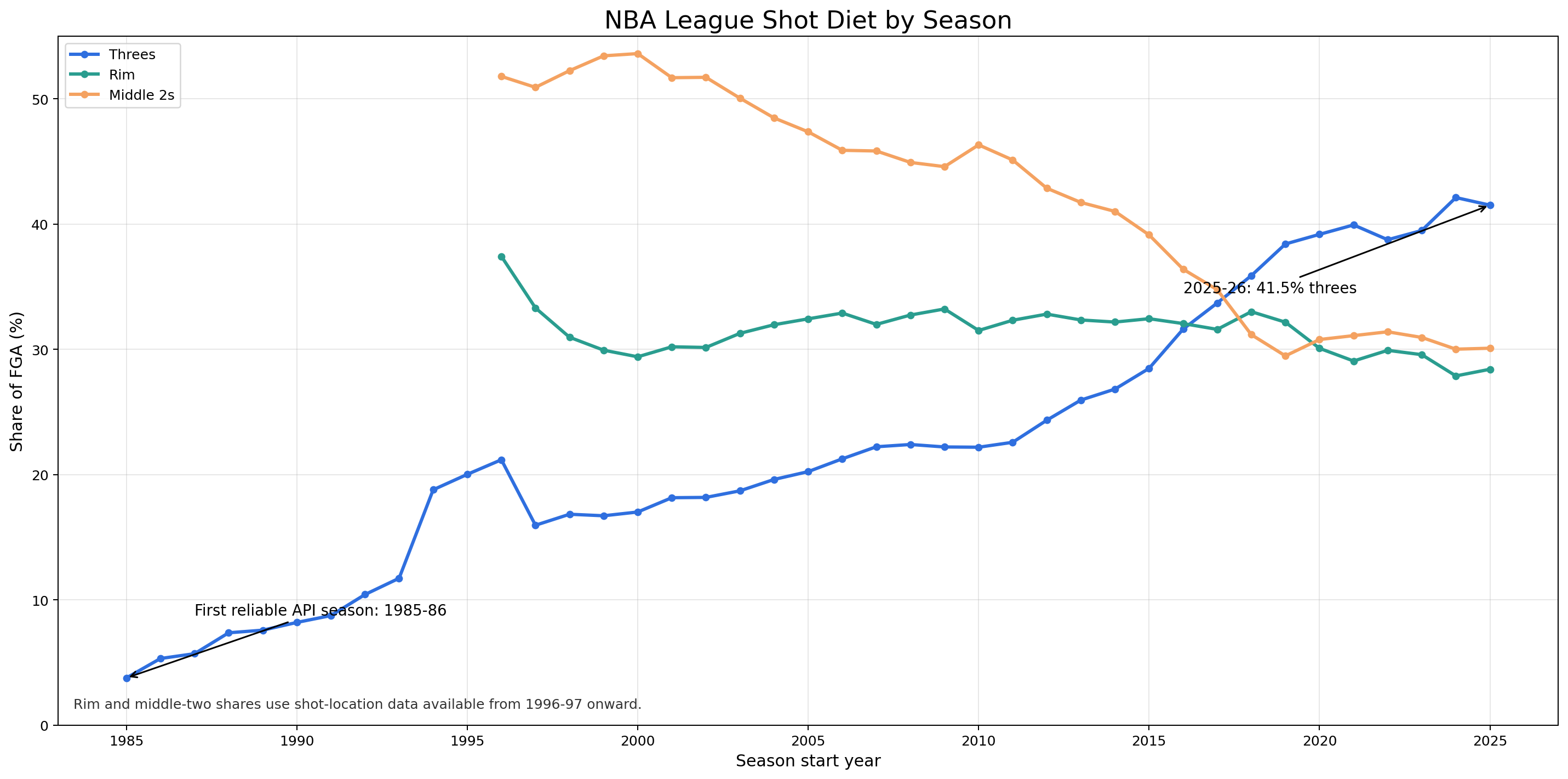}
\caption{League shot-diet share by season.  The three-point line shows the macro shift toward threes, while the rim and middle-two lines show how the rest of the shot diet moved around it.}
\label{fig:three-share}
\end{figure}

The basic expected-value argument is simple:
\[
E[V_z]=p_z r_z,
\]
where $p_z$ is make probability and $r_z$ is the point value of shot category $z$.  But expected value alone is incomplete.  Three-point shots can have strong expected value while still producing higher game-to-game variance.  Over the finite sample size of a single game or playoff series, a high-variance shot profile exposes a team to downside risk and cold streaks.
It also treats a miss as the same kind of event across shot types.  In reality, misses create different transition opportunities: some misses are more likely to become offensive rebounds, free throws, or another shot before the possession ends.

The abrupt shift in three-point shooting trends, alongside its current analytical justification, motivates the central question of this paper. Have NBA teams reached a point where they shoot too many threes? To investigate this, we develop an optimization model to benchmark empirical shot diets against the league-wide efficient frontier. Furthermore, we apply this framework to team-specific transition matrices. This allows us to generate customized shot portfolio recommendations for each franchise while validating directional consistency across unique offensive environments.
\begin{figure}[H]
\centering
\begin{minipage}{0.48\textwidth}
\centering
\includegraphics[width=\textwidth]{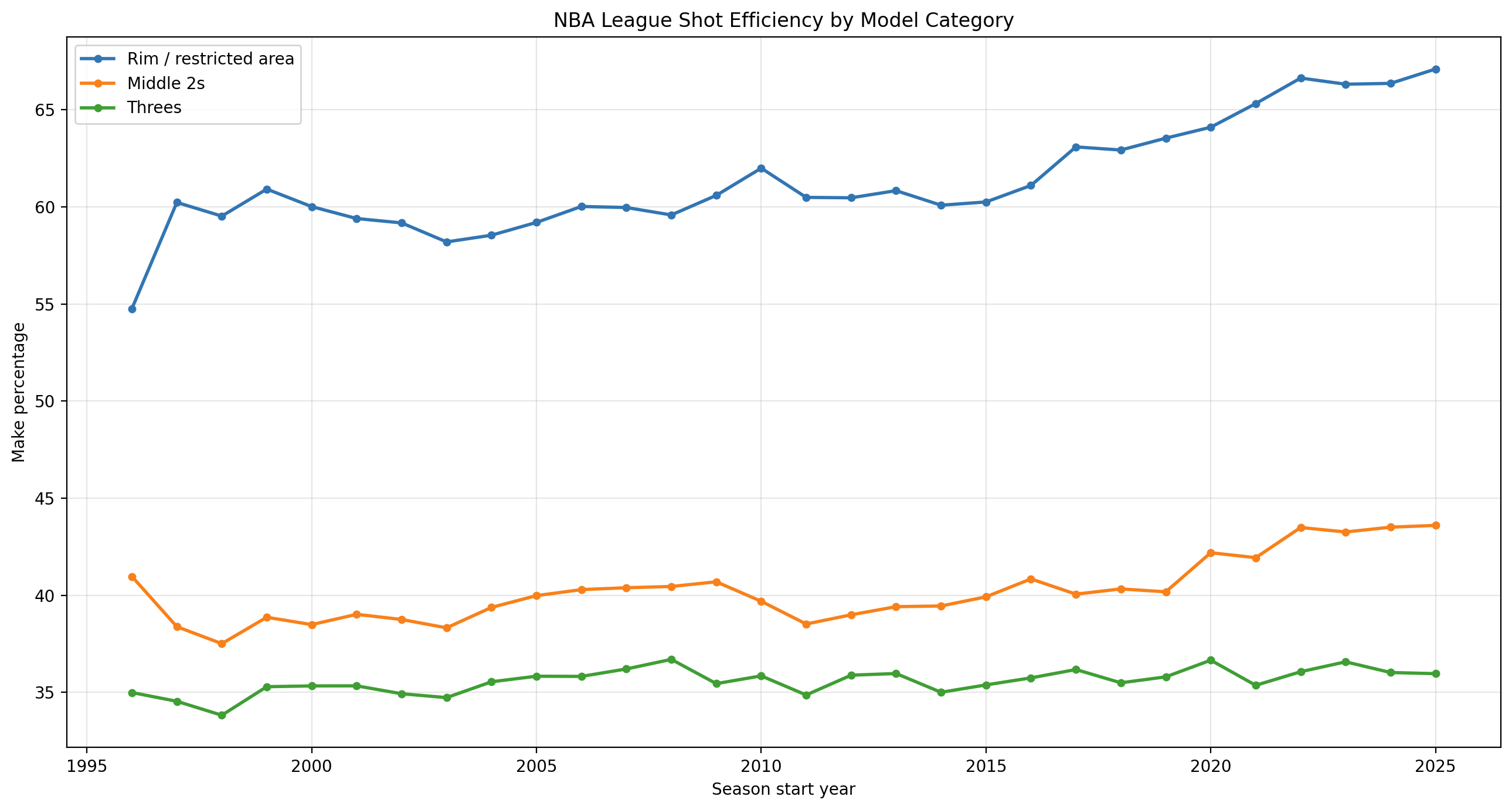}
\end{minipage}
\hfill
\begin{minipage}{0.48\textwidth}
\centering
\includegraphics[width=\textwidth]{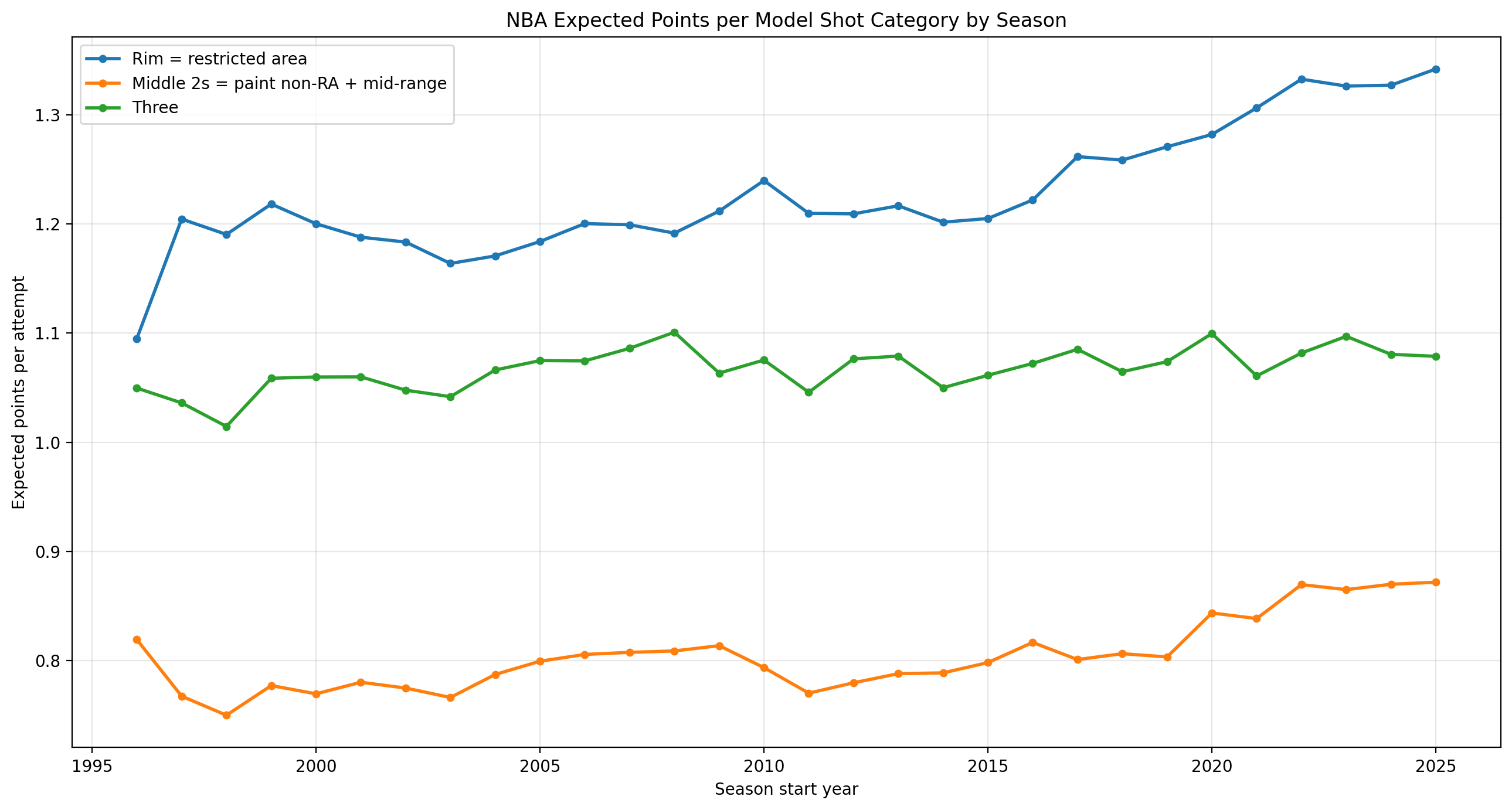}
\end{minipage}
\caption{Shot probability and shot value move differently by category.  Rim attempts have the strongest make-rate profile, while threes gain value from the extra point.  This is why raw expected value is necessary but not sufficient.}
\label{fig:historical-shot-values}
\end{figure}

This paper fills the gap between the micro and macro domains.  We model the possession as an absorbing Markov reward process, use the transient possession value as the return term in a portfolio objective, and penalize both shot-return variance and timeout-response instability.  The shot portfolio is
\[
\pi=(\pi_R,\pi_M,\pi_{C3},\pi_{NC3}),
\]
where $R$ is rim attempts, $M$ is midrange attempts, $C3$ is corner threes, and $NC3$ is non-corner threes.  The components satisfy
\[
\pi_R+\pi_M+\pi_{C3}+\pi_{NC3}=1,\qquad \pi_i\ge 0.
\]

We do not claim that one universal shot chart is correct. We will attempt to generalize the league-wide shot diets of the 3-point revolution and compare them to constructed efficient frontiers where continuation, variance, and response sensitivity are all included.

\section{Background and Gap}

Initial advancements in the basketball analytics revolution focused on the "micro" problem of offensive efficiency. Frameworks ranging from Skinner’s optimal stopping models \citep{skinner2011} to the Markov Decision Processes of Sandholtz and Bornn \citep{sandholtz2020} quantified the cascading probabilities of a single 24-second possession. Consequently, offenses were re-engineered to generate high-value spatial states, such as rim attempts and corner threes, by using dribble penetrations to force defensive rotational dilemmas \citep{pelechrinis2021}.

The 3-point revolution led to a new "macro" problem in basketball analytics. The macroscopic shift in league shot diets toward perimeter volume raised a systemic question: Is there a mathematical limit to this logic? Optimizing a full-game offensive architecture solely around expected value dangerously ignores outcome volatility. Because three-point shots inherently possess a lower raw make probability, relying exclusively on a high-variance perimeter profile exposes a team to severe downside risk over a finite sample. As demonstrated by Fichman and O’Brien’s application of Modern Portfolio Theory to basketball \citep{fichmanobrien2018}, elite offensive design requires balancing expected points against portfolio variance. They predicted that the three-point revolution would reach its equilibrium at about 38\% of field-goal attempts \citep{fichmanobrienold}. However, that number was reached in 2019, with the three-point attempt share since pushing further past their predicted equilibrium to 41.5\%.

Fichman and O’Brien treated shot diets as non-transient, not taking account for rebound continuations. To refine their framework, we account for the sequential continuations of live possessions. Therefore, a drive to the rim captures both its immediate expected value and the subsequent opportunities it generates.

Our model bridges the “micro” and “macro” domains by fusing the stochastic modeling of individual possessions with the macroscopic Modern Portfolio approach to optimize a comprehensive offensive profile. By defining the half-court offense as an absorbing Markov Reward Process, we calculate transient-state continuation values that inherently capture spatial transition probabilities. We then integrate this transient expected value into a centralized objective function alongside empirical shot variance and timeout-response penalties. Extracting the efficient frontier from this model mathematically balances the generation of high-value shots against the mitigation of systemic volatility.
Ultimately, this dynamic portfolio framework allows us to mathematically evaluate whether the NBA shoots "too many" threes. 

The contribution is therefore specific: this paper combines a possession-level Markov reward process with a shot-diet portfolio objective, separates corner threes from non-corner threes, and compares full-sample results with a no-fastbreak sample.  Fastbreaks are analyzed separately because their value depends on geometry, who gets back on defense, and how the offense fills lanes.  Standard play-by-play does not fully observe that information.

\section{Data and State Space}

The analysis uses cached NBA live play-by-play data from regular-season games in the 2021--22 through 2025--26 seasons.  Each possession is mapped into the state space
\[
\begin{aligned}
\mathcal S=\{&PS,RM,RX,MM,MX,C3M,C3X,NC3M,NC3X,\\
&FT1,FT2,FT3,OR_R,OR_M,OR_{C3},OR_{NC3},OR,EP\}.
\end{aligned}
\]
Here $PS$ is possession start and $EP$ is the absorbing end-of-possession state.  Made and missed shots are split by location and result.  Offensive rebounds are also split by the type of missed shot that created them.

In the no-fastbreak sample, after removing any modeled possession containing a live fastbreak qualifier, the diet was
\[
R/M/C3/NC3 = 0.313 / 0.277 / 0.105 / 0.305.
\]
The no-fastbreak run removed 129,695 modeled fastbreak possessions and 302,262 modeled events.  This is not a claim that fastbreaks are unimportant.  It is the opposite: fastbreak value depends heavily on geometry and player recovery, which are not fully captured in standard play-by-play.

The observed full-sample league shot diet was
\[
R/M/C3/NC3 = 0.345 / 0.252 / 0.103 / 0.301.
\]

\section{Model}

\subsection{Transition Matrix}

Let $P$ be the empirical transition matrix:
\[
P_{ij}=\Pr(X_{t+1}=j\mid X_t=i).
\]
The end state is absorbing:
\[
P_{EP,EP}=1.
\]
For transient analysis, remove $EP$ and write the remaining transient block as $Q\equiv Q_{\text{trans}}$.  The fundamental matrix is
\[
N=(I-Q)^{-1}=I+Q+Q^2+Q^3+\cdots.
\]
The entry $N_{ij}$ is the expected number of visits to transient state $j$ before the possession ends when starting in state $i$.

For a candidate shot diet $\pi$, the transition matrix becomes $P(\pi)$.   $P(\pi)$ reallocates shot-outcome mass in the possession-start and offensive-rebound rows across the four shot categories while preserving empirical non-shot mass, free-throw mass, end-of-possession mass, and category make rates.  If $p_z$ is the make rate of category $z$ and $\omega$ is the row's field-goal outcome mass, then the simplified row logic is
\[
\Pr(\text{row}\rightarrow z_M\mid \pi)=\omega \pi_z p_z,\qquad
\Pr(\text{row}\rightarrow z_X\mid \pi)=\omega \pi_z(1-p_z).
\]
This is why $\pi$ changes the shot mix without pretending that turnovers, fouls, and possession-ending events disappear.

\begin{figure}[H]
\centering
\includegraphics[width=\textwidth]{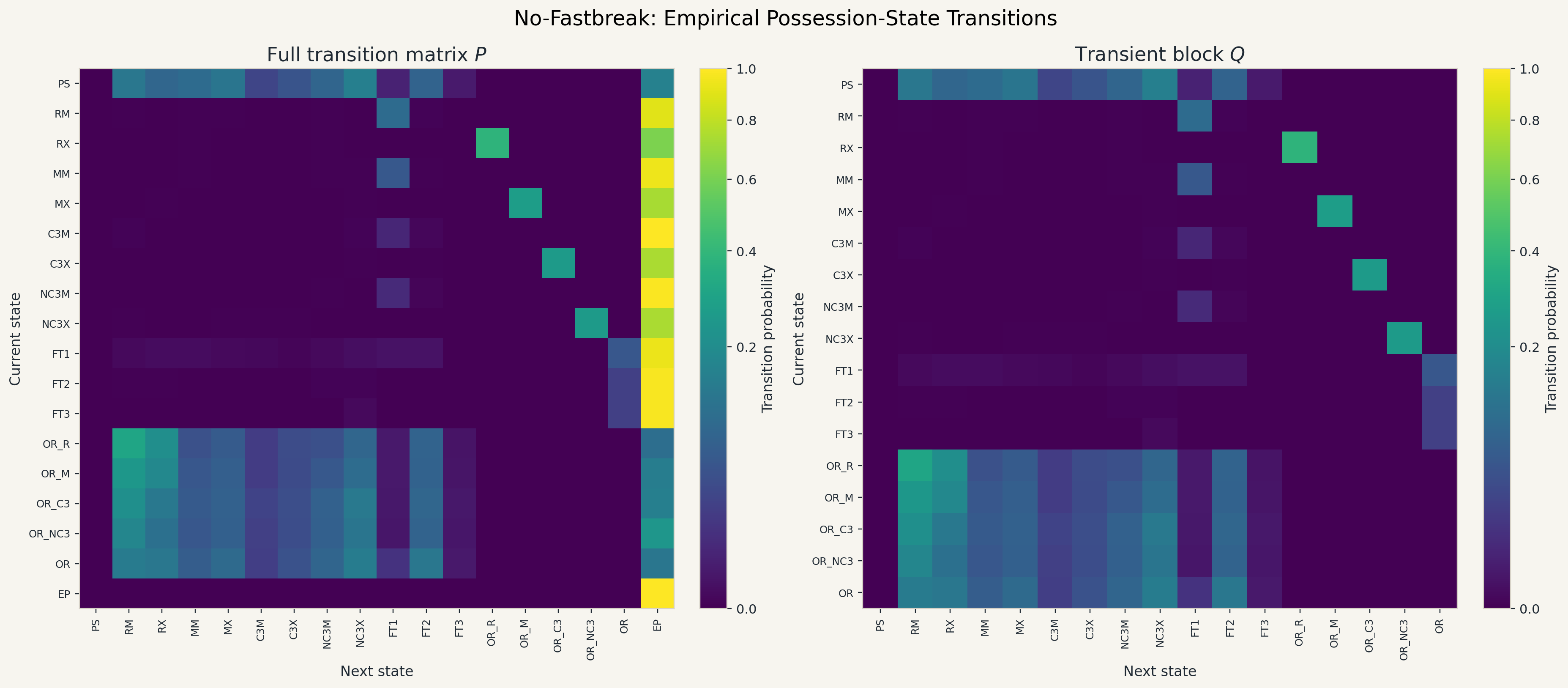}
\caption{No-fastbreak empirical transition matrices.  The left panel is the full transition matrix $P$ and the right panel is the transient block $Q_{\text{trans}}$ after removing the absorbing end-possession state.}
\label{fig:nf-pq}
\end{figure}

\begin{figure}[H]
\centering
\includegraphics[width=\textwidth]{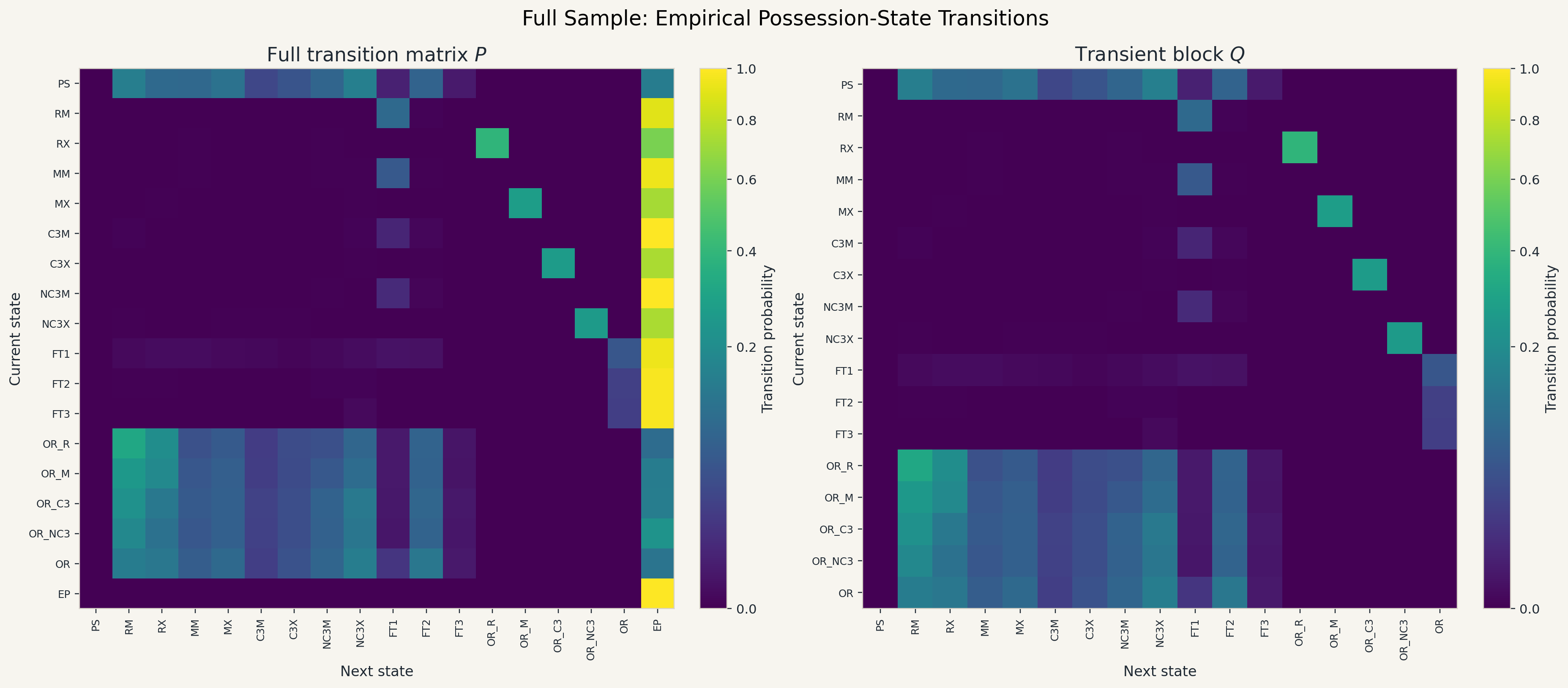}
\caption{Full-sample empirical transition matrices.  Adding fastbreaks increases the possession-start mass into rim outcomes, while the transient block still shows the same continuation channels through missed shots, free throws, and offensive rebounds.}
\label{fig:full-pq}
\end{figure}

Figures \ref{fig:nf-pq} and \ref{fig:full-pq} show the empirical transition structure.  The full matrix $P$ includes the absorbing $EP$ state, so it shows where possessions terminate.  The transient block $Q_{\text{trans}}$ removes $EP$ and isolates the continuation process before possession end.  The important rows for this paper are the missed-shot rows, because they determine whether a miss is truly a dead possession or a continuation opportunity.

The miss-to-offensive-rebound rows show that rim misses create the most continuation.  In the no-fastbreak sample,
\[
\begin{aligned}
\Pr(OR_R\mid R_X)&=0.384, & \Pr(OR_M\mid M_X)&=0.270,\\
\Pr(OR_{C3}\mid C3_X)&=0.258, & \Pr(OR_{NC3}\mid NC3_X)&=0.260.
\end{aligned}
\]
In the full sample the same ordering holds:
\[
\begin{aligned}
\Pr(OR_R\mid R_X)&=0.391, & \Pr(OR_M\mid M_X)&=0.271,\\
\Pr(OR_{C3}\mid C3_X)&=0.262, & \Pr(OR_{NC3}\mid NC3_X)&=0.260.
\end{aligned}
\]
So, in this model, missed rim attempts have the highest offensive-rebound continuation opportunity.  Corner-three and non-corner-three misses are nearly tied, and both are slightly below midrange misses.

\begin{figure}[H]
\centering
\includegraphics[width=0.82\textwidth]{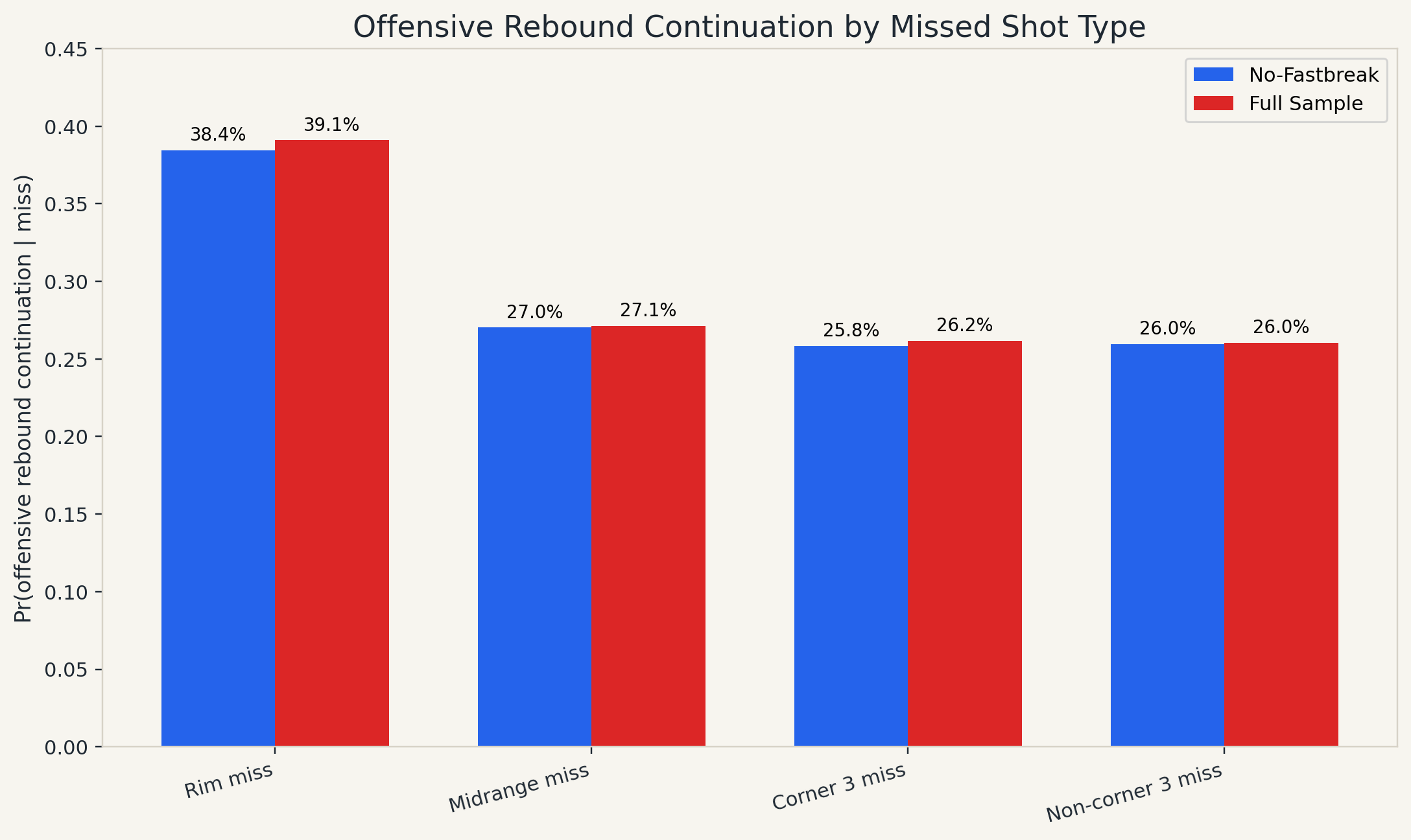}
\caption{Offensive-rebound continuation probability conditional on a missed shot.  Rim misses are the strongest continuation source in both samples.}
\label{fig:or-continuation-rates}
\end{figure}

\subsection{Rewards, One-Step Value, and Transient Value}

The reward vector $r$ attaches immediate points to entering states:
\[
r_{RM}=2,\qquad r_{MM}=2,\qquad r_{C3M}=3,\qquad r_{NC3M}=3.
\]
Free-throw states receive their empirical expected free-throw contribution, and miss/end states receive zero immediate reward.

The one-step value is
\[
V_1(\pi)=e_{PS}^{T}Q(\pi)r.
\]
This is the expected value of the next transition from possession start.  It is useful because it is close to the traditional first-shot expected-value view.

The transient value is
\[
V_T(\pi)=e_{PS}^{T}(I-Q(\pi))^{-1}r=e_{PS}^{T}N(\pi)r.
\]
This is the expected value of the possession before absorption into $EP$.  It is the main value object because it includes continuation after offensive rebounds, free throws, and possession survival.  Comparing $V_1$ and $V_T$ shows exactly where continuation changes the ranking of shot diets.

\subsection{Variance}

Let $g_t$ be the game-level vector of shot-category returns:
\[
g_t = [g_{t,R},g_{t,M},g_{t,C3},g_{t,NC3}]^T,
\qquad
g_{t,z}=\frac{\text{points from category }z\text{ in game }t}
{\text{attempts from category }z\text{ in game }t}.
\]
The covariance matrix is
\[
\Sigma=\frac{1}{T-1}\sum_{t=1}^{T}(g_t-\bar g)(g_t-\bar g)^T.
\]
For a candidate diet, portfolio variance is
\[
\sigma^2(\pi)=\pi^T\Sigma\pi
=\sum_i\sum_j \pi_i\Sigma_{ij}\pi_j.
\]
We use covariance rather than independent category variance because shot families can co-move.  A high-rim game and a high-corner-three game may not be independent events; covariance is the part of the model that lets diversification matter.

\subsection{Timeout Response Cost}

Timeout intervals are split on team-attributed full timeouts.  Each interval has a pressure vector
\[
s_t=[EP_R,EP_M,EP_{C3},EP_{NC3},EP_{OR},EP_{FT}]^T.
\]
The first four components are interval scoring pressure by shot family:
\[
EP_{z,t}=\text{share}_{z,t}\cdot \text{FG\%}_{z,t}\cdot r_z.
\]
The remaining components are
\[
\begin{aligned}
EP_{FT,t} &=
\frac{\text{free-throw points in interval }t}{FGA_t},\\
EP_{OR,t} &=
\frac{\text{points after first offensive rebound in interval }t}{FGA_t}.
\end{aligned}
\]

The response matrix is fit by ridge regression:
\[
\Delta s_t = B s_t+\epsilon_t,\qquad
\hat B^T=(X^TX+\alpha I)^{-1}X^TY,
\]
where each row of $X$ is the current interval pressure vector and each row of $Y$ is the next-interval change $s_{t+1}-s_t$.

For a candidate diet, the model builds
\[
s(\pi)=[s_R(\pi),s_M(\pi),s_{C3}(\pi),s_{NC3}(\pi),s_{OR}(\pi),s_{FT}(\pi)]^T,
\]
with
\[
s_z(\pi)=\Pr(PS\rightarrow z_M\mid \pi)r_z,
\]
\[
s_{OR}(\pi)=\max\{V_T(\pi)-V_1(\pi),0\},
\qquad
s_{FT}(\pi)=\sum_{k\in\{FT1,FT2,FT3\}}\Pr(PS\rightarrow k\mid \pi)r_k.
\]
The response cost is
\[
E(\pi)=\|B s(\pi)\|_2.
\]
We include $B$ because a shot diet can be efficient but structurally unstable. This term functions as an implicit coaching adjustment cost, penalizing shot profiles that empirically trigger significant post-timeout movement. Practically, if an offense generates an unsustainably high volume of uncontested three-pointers, the opposing defense will use the timeout to structurally regress that specific advantage back toward the mean.

\subsection{Optimization and Frontier Selection}

The final objective is
\[
\max_\pi\ \widetilde V(\pi)
-\lambda_1\widetilde{\sigma^2}(\pi)
-\lambda_2\widetilde E(\pi),
\]
subject to
\[
0\leq \pi_i\leq 1,\qquad \sum_i\pi_i=1.
\]
Each term is normalized on the candidate portfolio support before entering the objective:
\[
\widetilde A(\pi)=\frac{A(\pi)-A_{\min}}{A_{\max}-A_{\min}}.
\]
This keeps value, variance, and response cost on comparable scales.  Without this normalization, the largest numerical term can dominate the objective for a mechanical reason rather than a basketball reason.
A grid search across $\lambda_1$ and $\lambda_2$ is used to trace the frontier and identify the risk-return region where the knee rule is applied.  The league selected portfolios with the respective  $\lambda_1$ and $\lambda_2$  are then solved with constrained SLSQP over the continuous simplex. 

The main knee rule in this paper is the dx/dy-tail rule.  Let the upper efficient envelope be ordered as
\[
(x_m,y_m)=\left(\sigma^2(\pi_m),V(\pi_m)\right).
\]
The rule computes the local tail slope
\[
\left|\frac{dx}{dy}\right|_m
\approx
\left|\frac{x_{m+1}-x_m}{y_{m+1}-y_m}\right|
\]
and selects the high-value frontier point where this quantity is smallest.  Intuitively, this is the part of the frontier where the curve has entered its nearly vertical tail: more risk is being added faster than useful value.  We use it as the main rule because it is conservative and matches the question of risk-controlled shot allocation.

\subsection{Team-Level Matrices}

League optimization is a benchmark, but team-specific optimization needs team-level transition matrices.  For the main team figures, we estimate each team's transition matrix directly from its observed transition counts.  The five-season sample gives each team a reasonable number of modeled transitions, including more than 60,000 no-fastbreak training transitions for every team, so the main analysis does not need an added smoothing term.  This keeps the team-specific results tied to each team's own observed possession structure.

\subsection{Hybrid Fastbreak Recomposition}

The main optimizer excludes fastbreak possessions because their value depends on geometry that standard play-by-play does not observe.  To still express a full-game recommendation, we add fastbreaks back as an observed fixed component rather than optimizing them.  Let $N^{FB}$ be the number of modeled fastbreak possessions and $N^{All}$ be the total number of modeled possessions.  The fastbreak weight is
\[
\alpha=\frac{N^{FB}}{N^{All}}.
\]

For shot category $z$, let $A^{FB}_z$ be attempts from that category inside the removed fastbreak possessions only.  The observed fastbreak shot distribution is
\[
\rho^{FB,L}_z=\frac{A^{FB}_z}{\sum_j A^{FB}_j},
\]
The hybrid recommendation uses the no-fastbreak optimum and fastbreak distribution:
\[
\pi^{Hybrid}_z
=(1-\alpha)\pi^{NF*}_z+\alpha\rho^{FB}_z.
\]
In the data, the league fastbreak-possession probability is $\alpha^L=0.105$.  The league fastbreak shot distribution is
\[
R/M/C3/NC3 = 0.571 / 0.070 / 0.090 / 0.269.
\]
This is why adding the observed fastbreak component raises rim share and lowers midrange share relative to the no-fastbreak optimum.

\section{Results}
\subsection{Response Cost}
\begin{figure}[H]
\centering
\includegraphics[width=0.78\textwidth]{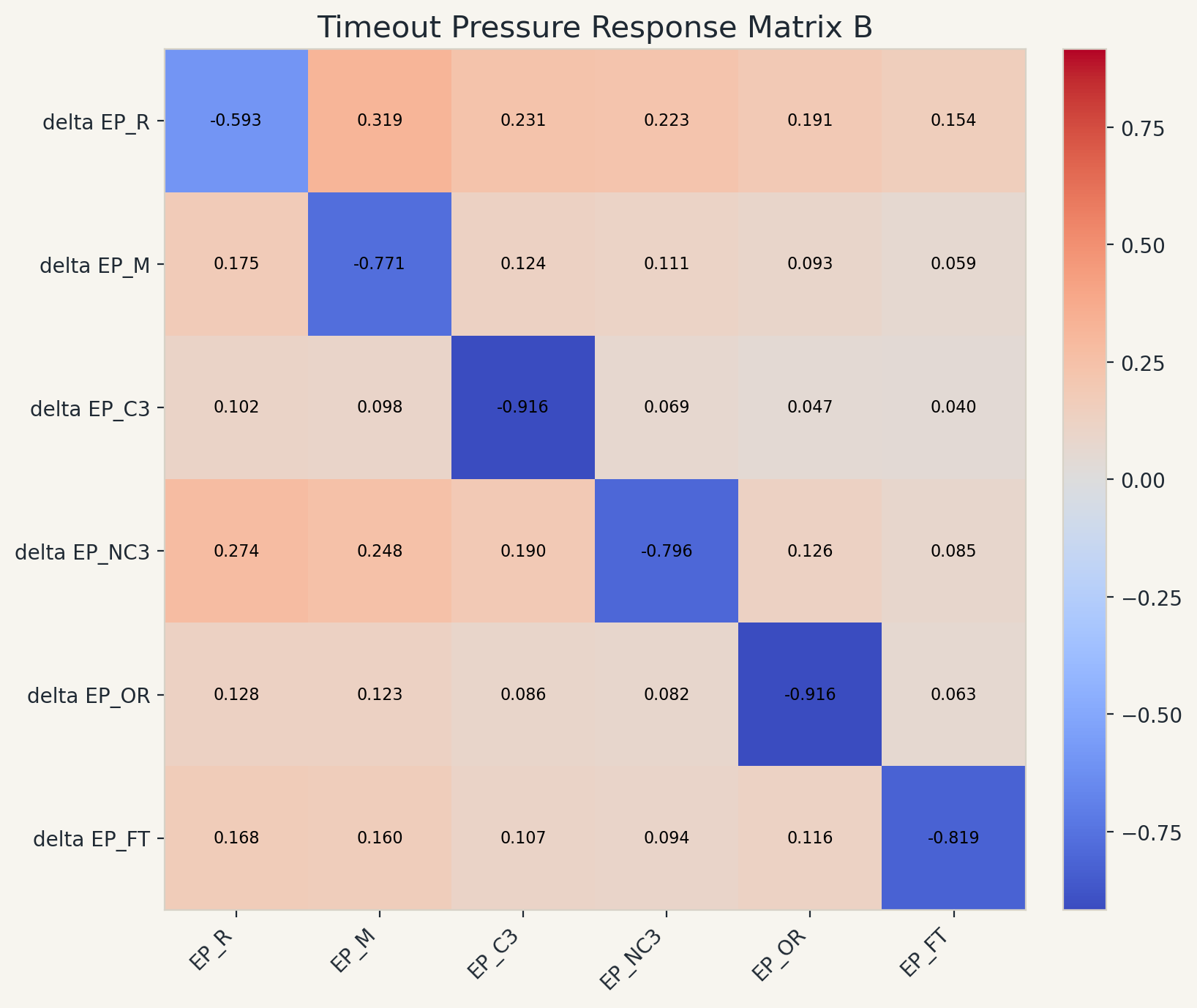}
\caption{Estimated timeout response matrix $B$ for the six pressure components.  Negative diagonal entries indicate mean reversion within the same pressure channel, while off-diagonal entries show how one pressure source is associated with movement in another.}
\label{fig:response-matrix-b}
\end{figure}

Using timeout-separated pressure intervals, the response cost matrix $B$ can be seen in Figure \ref{fig:response-matrix-b}.  This response cost takes into account the ability for the defense to quickly adapt to the shooting style of the opposing team.  To figure out what type of offense minimizes $\|B s(\pi)\|_2$ we want to find a $\pi$ where:

\[
\|B s(\pi) - s(\pi)\|_2 \approx 0 
\]
In Table \ref{tab:response-cost-minimum}, it can be seen that the $\pi$ that minimizes $\|B s(\pi)\|_2$ is approximately the actual league shooting distribution. This means that the response cost yields approximately the same minimum as a standard $L_2$ penalty, an expected result given its derivation from aggregated empirical data. However, matrix $B$ ensures that our penalty behaves asymmetrically, accurately capturing the cross-correlated cost of different spatial adjustments.

\begin{table}[H]
\centering
\caption{Response-cost-minimizing shot diet compared with the full-sample actual league diet.}
\label{tab:response-cost-minimum}
\begin{tabular}{lrrrr}
\toprule
Diet & Rim & Mid & C3 & NC3 \\
\midrule
Full actual & 0.345 & 0.252 & 0.103 & 0.301 \\
Minimum response cost & 0.342 & 0.253 & 0.108 & 0.297 \\
\bottomrule
\end{tabular}
\end{table}

\subsection{League-Level Shot Diets}

Table \ref{tab:league-optima} reports the no-fastbreak league results under the normalized w-tail objective and the hybrid full-game recomposition.  It can be seen that $\lambda_1$ and $\lambda_2$ are similar across the one-step and transient models.
\begin{table}[H]
\centering
\caption{League optima under dx/dy-tail frontier selection.}
\label{tab:league-optima}
\begin{tabular}{llrrrrrrr}
\toprule
Sample & Model & $\lambda_1$ & $\lambda_2$ & Rim & Mid & C3 & NC3 & Value \\
\midrule
No fastbreak & One-step & 5.8 & 0.7 & 0.404 & 0.230 & 0.100 & 0.266 & 0.945 \\
No fastbreak & Transient & 5.7 & 0.7 & 0.415 & 0.233 & 0.095 & 0.257 & 1.099 \\
Full Sample & Transient & 5.7 & 0.7 & 0.431 & 0.216 & 0.095 & 0.258 & -- \\
\bottomrule
\end{tabular}
\end{table}

Fastbreaks are valuable but structurally different from half-court possessions.  Their expected value depends on where players are on the floor, how many defenders get back, and how quickly the offense fills lanes.  The standard play-by-play API contains a fastbreak qualifier, but it does not contain the full geometry.  For this reason, the no-fastbreak version removes any modeled possession containing a fastbreak-tagged event.

The no-fastbreak actual diet has fewer rim attempts and more midrange attempts than the full sample.  Table \ref{tab:actual-shot-diets-by-sample} separates the half-court-heavy possessions from the removed fastbreak possessions.

\begin{figure}[H]
\centering
\includegraphics[width=\textwidth]{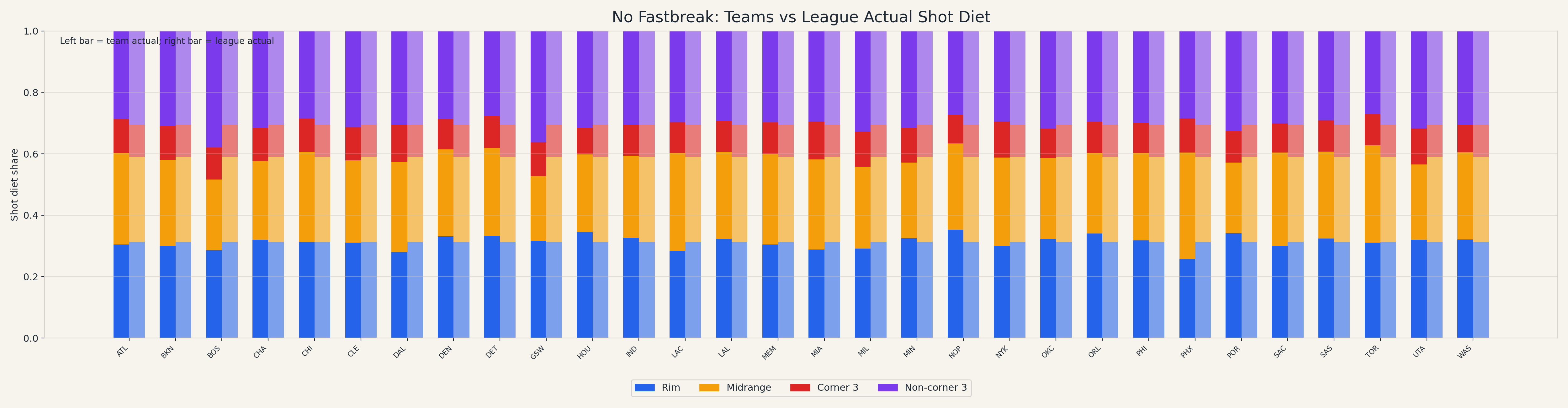}
\caption{No-fastbreak team diets compared with the no-fastbreak league actual diet.}
\label{fig:nf-actual}
\end{figure}

\begin{figure}[H]
\centering
\includegraphics[width=\textwidth]{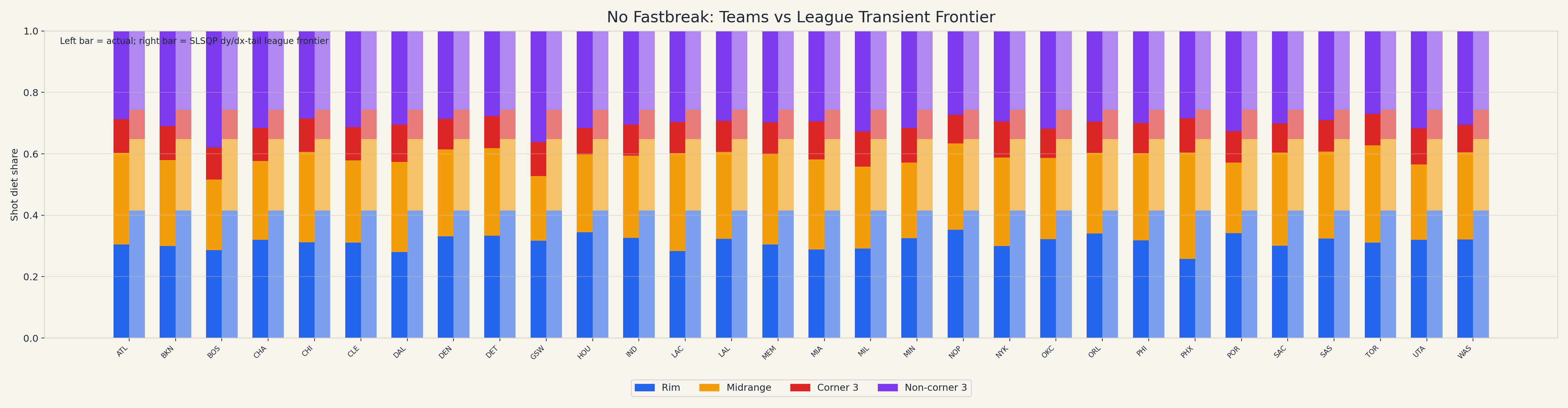}
\caption{No-fastbreak team diets compared with the no-fastbreak league transient frontier.}
\label{fig:nf-frontier}
\end{figure}

\begin{table}[H]
\centering
\caption{Actual league shot diets by possession sample.}
\label{tab:actual-shot-diets-by-sample}
\begin{tabular}{lrrrr}
\toprule
Sample & Rim & Mid & C3 & NC3 \\
\midrule
No fastbreak actual & 0.313 & 0.277 & 0.105 & 0.305 \\
Fastbreak actual & 0.571 & 0.070 & 0.090 & 0.269 \\
Full sample actual & 0.345 & 0.252 & 0.103 & 0.301 \\
\bottomrule
\end{tabular}
\end{table}
Figure \ref{fig:nf-actual} and \ref{fig:nf-frontier} show each teams actual shot diet compared to the league actual and optimum respectfully. The no-fastbreak transient optimum moves meaningfully toward rim pressure while staying diversified:

\begin{table}[H]
\centering
\caption{No-fastbreak actual diet compared with the no-fastbreak transient optimum.}
\label{tab:no-fastbreak-actual-vs-optimum}
\begin{tabular}{lrrrr}
\toprule
Diet & Rim & Mid & C3 & NC3 \\
\midrule
No fastbreak actual & 0.313 & 0.277 & 0.105 & 0.305 \\
No fastbreak transient optimum & 0.415 & 0.233 & 0.095 & 0.257 \\
\bottomrule
\end{tabular}
\end{table}
The recommendation is more rim pressure and fewer non-corner threes than the observed diet, but it is not a pure rim-and-three profile.

The hybrid full-game benchmark uses the no-fastbreak transient optimum for half-court-like possessions and the observed fastbreak shot distribution for transition possessions:

\begin{table}[H]
\centering
\caption{Hybrid actual diet compared with the Hybrid transient optimum.}
\label{tab:hybrid-actual-vs-optimum}
\begin{tabular}{lrrrr}
\toprule
Diet & Rim & Mid & C3 & NC3 \\
\midrule
Hybrid actual & 0.340 & 0.255 & 0.103 & 0.301 \\
Hybrid transient optimum & 0.431 & 0.216 & 0.095 & 0.258 \\
\bottomrule
\end{tabular}
\end{table}
This is directionally consistent with transition offense creating easier rim chances.  The result remains similar because only the non-fastbreak component is optimized; the fastbreak component is held to the actual removed fastbreak shot distribution.

\begin{figure}[H]
\centering
\includegraphics[width=\textwidth]{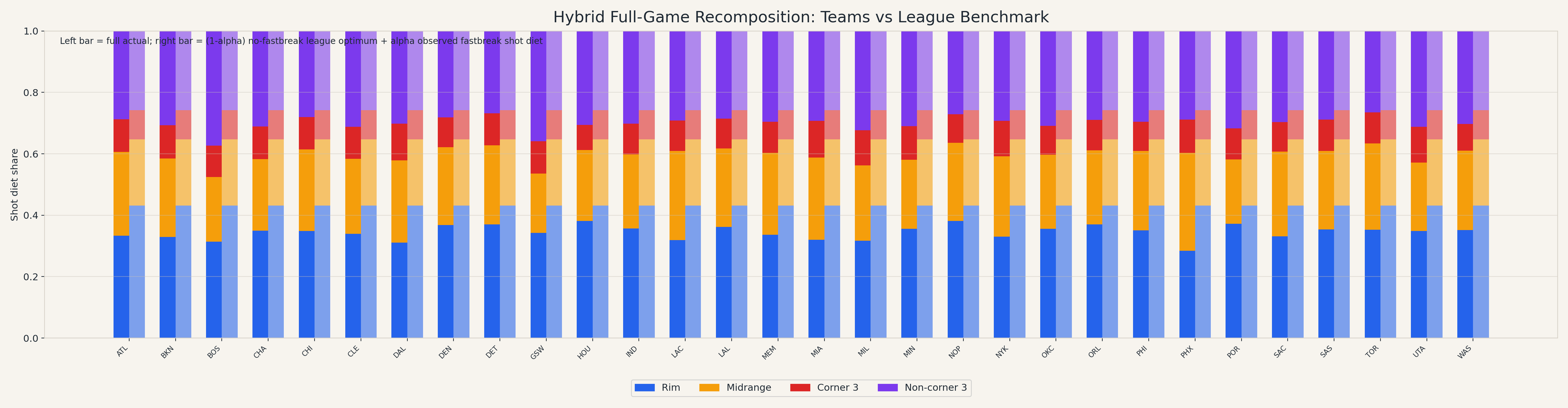}
\caption{Full actual team diets compared with the hybrid league benchmark.  The right-hand bar is $(1-\alpha^L)$ times the no-fastbreak league transient optimum plus $\alpha^L$ times the league fastbreak shot distribution.}
\label{fig:hybrid-league-bars}
\end{figure}
\begin{figure}[H]
\centering
\includegraphics[width=0.88\textwidth]{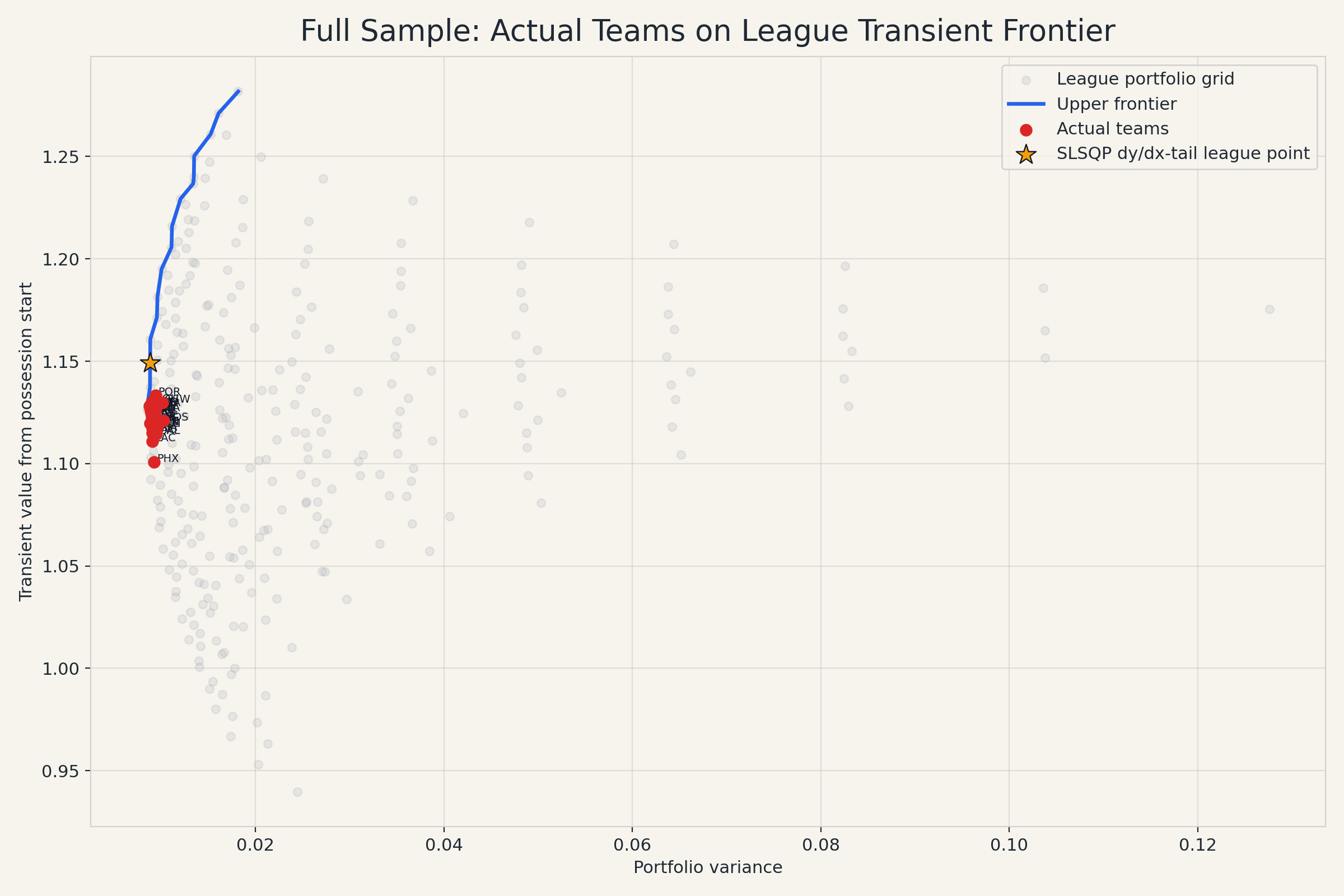}
\caption{Actual teams plotted on the full-sample league transient frontier.  The dx/dy-tail point is the selected league frontier portfolio when all possessions, including fastbreaks, are optimized together.}
\label{fig:full-transient-frontier}
\end{figure}

As a calibration check, Figure \ref{fig:full-transient-frontier} shows the full-sample league transient frontier when fastbreak and non-fastbreak possessions are pooled.  The selected full-sample model value is 1.149 points per possession (PPP), close to the empirical league PPP of about 1.15, so the Markov reward scale is aligned with observed scoring.

In Figure 11, we see that the empirical shot diets of all NBA franchises live in close proximity to the league-wide efficient frontier. This indicates that modern offensive distributions are already highly optimized within our framework. In Figure 10, see that marginal improvements can be made by slightly reducing their three-point and midrange attempt shares while proportionally increasing rim volume. 
\subsection{Team-Specific Frontiers}

In this section, we examine what the model recommends under teams' own statistical environments.  Teams differ in shot efficiency, offensive rebounding paths, covariance structure, and transition tendencies.  The team-specific results support the main conceptual point: there is not one universal shot diet.

\begin{figure}[H]
\centering
\includegraphics[width=\textwidth]{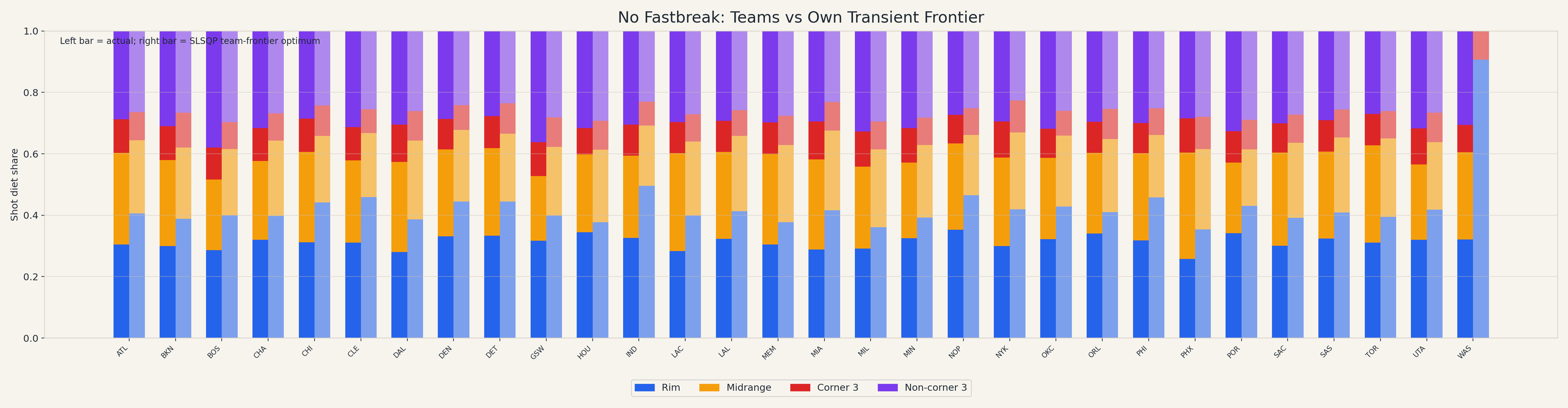}
\caption{No-fastbreak team shot diets compared with team-specific transient frontier portfolios.}
\label{fig:nf-team-bars}
\end{figure}

\begin{figure}[H]
\centering
\includegraphics[width=\textwidth]{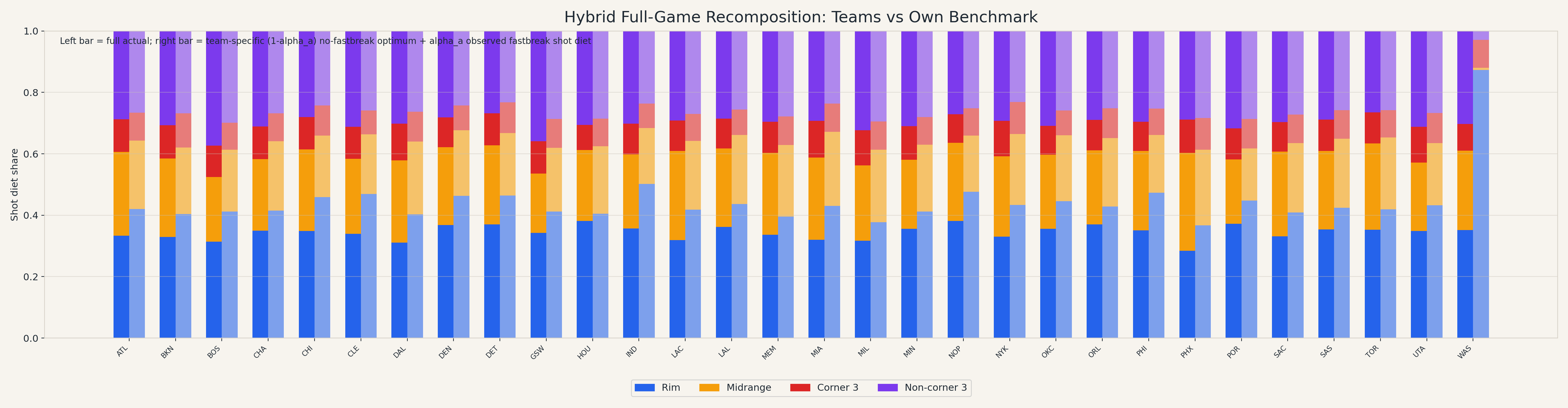}
\caption{Full actual team diets compared with team-specific hybrid benchmarks.  The right-hand bar uses each team's own no-fastbreak optimum, fastbreak-possession probability, and observed fastbreak shot distribution.}
\label{fig:hybrid-team-bars}
\end{figure}

For the team-specific comparison figures, the selected team points use each team's empirical transition matrix. The resulting target distributions exhibit the identical structural trends observed along the macro league-wide frontier. As illustrated in Figures \ref{fig:nf-team-bars} and \ref{fig:hybrid-team-bars}, our model suggests that the majority of teams should shoot more rim shots, a similar amount of corner threes, and fewer non-corner threes and midrange shots.

We observe that Washington's optimum represents an extreme anomaly, suggesting a portfolio composed almost entirely of rim attempts. To address this, we detail an empirical shrinkage approach in the appendix. While shrinkage successfully corrected Washington's benchmark and maintained stability across the majority of the league, it simultaneously introduced severe volatility into Denver's optimum. We hypothesize that this divergence highlights a fundamental limitation of the shrinkage estimator: it correctly regresses Washington's small-sample noise to the mean but inappropriately penalizes Denver's statistically extreme, yet highly efficient, structural reality.
\section{Discussion}

The main conclusion is that the three-point revolution should not be evaluated as a slogan. The overarching problem is not whether a three-point attempt is inherently worth more than a two-point jump shot. Rather, the question is what specific shot portfolio maximizes transient possession value while simultaneously controlling for outcome volatility and strategic instability. Under our model, NBA offenses are already operating very close to the efficient frontier, leaving only marginal gains to be extracted. Specifically, the model's directional optimum supports a slight reduction in perimeter and midrange volume in favor of a corresponding increase in rim attempts.

The no-fastbreak results are important because they keep the model honest.  Transition offense is generated by spacing, live-ball turnovers, rebound outlets, player speed, and defensive recovery.  Without tracking data, we cannot fully observe those geometric inputs.  Removing fastbreak possessions therefore gives a cleaner view of half-court shot allocation.

Our results demonstrate that a single, universally optimal shot diet does not exist, as teams have unique suggestions based on their statistical landscapes. However, the mathematical divergence between these team-specific optimums is generally narrower than the stark stylistic contrasts observed in the contemporary NBA. For instance, the Denver Nuggets and Boston Celtics maintain historically elite offensive efficiencies despite utilizing opposite perimeter-volume philosophies. This indicates that while the underlying statistical landscape defines the structural boundaries of a team's efficiency, unique personnel constraints and overarching coaching philosophy dictate its ultimate behavioral extreme. While our $dx/dy$ stopping condition identifies a mathematical equilibrium between risk and reward, individual franchises may weigh risk and reward differently based on their personnel. A system built around high-volume perimeter creation and one built around interior pressure can both be mathematically coherent. This does not mean all shot diets are created equal; it proves that true offensive optimality is conditional on both a team's statistical ecosystem and its unique personnel and offensive philosophy.

\section{Limitations and Future Work}

The model holds the empirical make percentage constant when shot shares change.  We deliberately do this to avoid overfitting.  A simple regression of shot attempts on shooting percentage would be flawed because attempts are not randomly assigned: a team may shoot more from a category because it is hot that day, because the defense is conceding that shot, or because personnel created unusually good looks.  Further work with optical tracking data is required to estimate how efficiency changes when shot frequency changes.

The framework functions most effectively as a macro-level prescription rather than a granular optimization tool for individual franchises. On a localized, team-specific level, the model lacks critical personnel identification. We do not explicitly quantify the offensive gravity of distinct superstars, specific lineup spacing, matchup pressure, or individual shot-maker identities. Because the model smooths over these localized variables, the team-specific frontiers it generates naturally mirror the macro league optimum. Therefore, while our overarching directional recommendation—increasing rim pressure while reducing midrange and perimeter volume—serves as a highly robust baseline for the average franchise, it should not be viewed as an absolute mandate. In reality, a team's true efficient frontier would dynamically shift based on its specific personnel advantages, meaning this macro directional trend may not hold true for every individual roster.

Fastbreaks are another limitation.  We exclude them in the main no-fastbreak analysis because transition offense depends on geometry: who gets back, who fills lanes, where the ball is, and how the defense is matched.  A stronger model would estimate fastbreak creation and fastbreak shot diet as explicit optimizable terms rather than treating transition possessions separately.

Finally, we do not explicitly model defense.  The timeout response matrix is only an implicit way to measure adaptability after pressure states.  Proper defensive modeling would require more advanced data: exact coverage responses, help positioning, matchup changes, substitutions, and personnel-specific defensive choices.

While this framework identifies the macro-level efficient frontier for expected possession value, future research should map these portfolios directly to expected win probability. The central question for subsequent models is whether shifting the objective function from expected points to expected wins naturally bridges the slight gap between our theoretical optimum and observed league averages. By optimizing the league-wide shot diet against game-state simulators, future work can definitively determine if the modern NBA's current, high-variance empirical shot distribution is actually the mathematical requirement for win maximization.

\appendix

\section{Full-Sample Frontier Figures}

The main text uses the no-fastbreak optimizer plus observed fastbreak recomposition.  For comparison, the original full-sample SLSQP figures are shown here.  These figures optimize directly on the full sample, so fastbreak and non-fastbreak possessions are pooled in a single transition matrix.

\begin{figure}[H]
\centering
\includegraphics[width=0.88\textwidth]{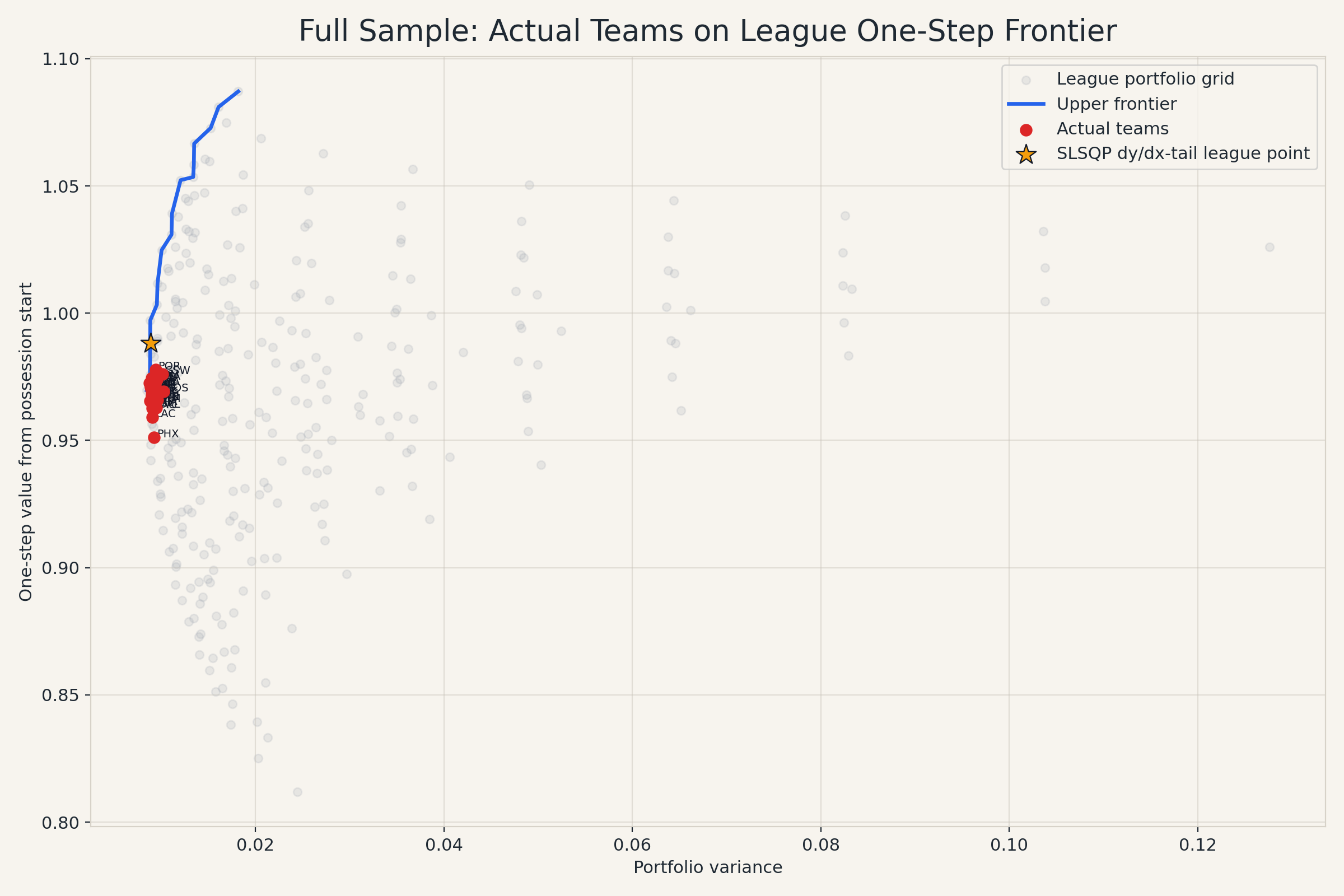}
\caption{Actual teams plotted on the full-sample league one-step frontier.  The one-step model shows the first-transition value surface before full continuation is included.}
\label{fig:appendix-full-one-step-frontier}
\end{figure}

\begin{figure}[H]
\centering
\includegraphics[width=\textwidth]{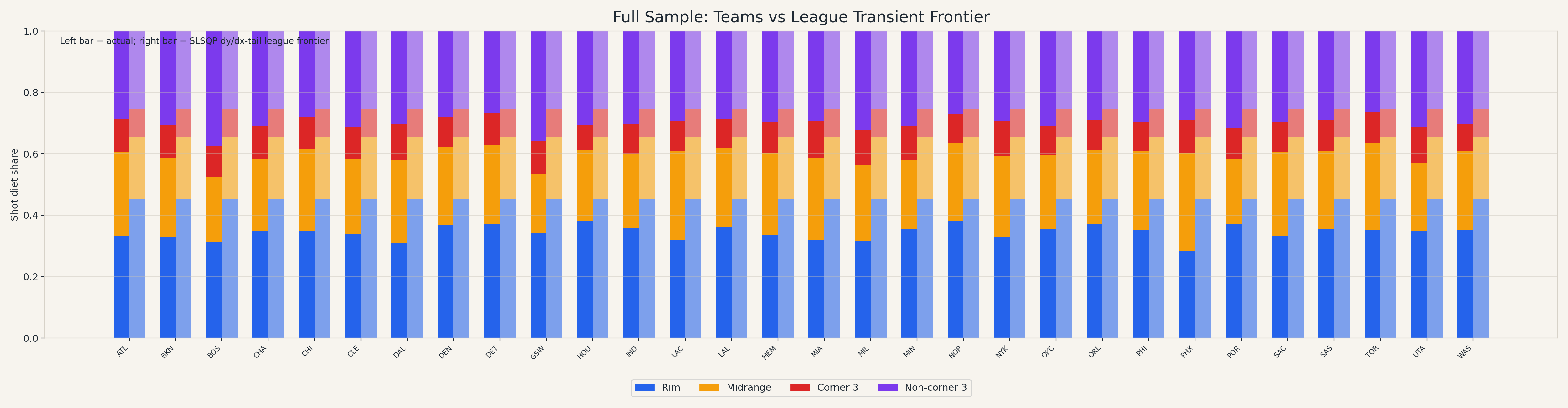}
\caption{Full-sample team shot diets compared with the league transient frontier portfolio selected by dx/dy tail.  This is a full-sample optimization over all possessions, so fastbreaks are not held fixed.}
\label{fig:appendix-full-league-bars}
\end{figure}

\begin{figure}[H]
\centering
\includegraphics[width=\textwidth]{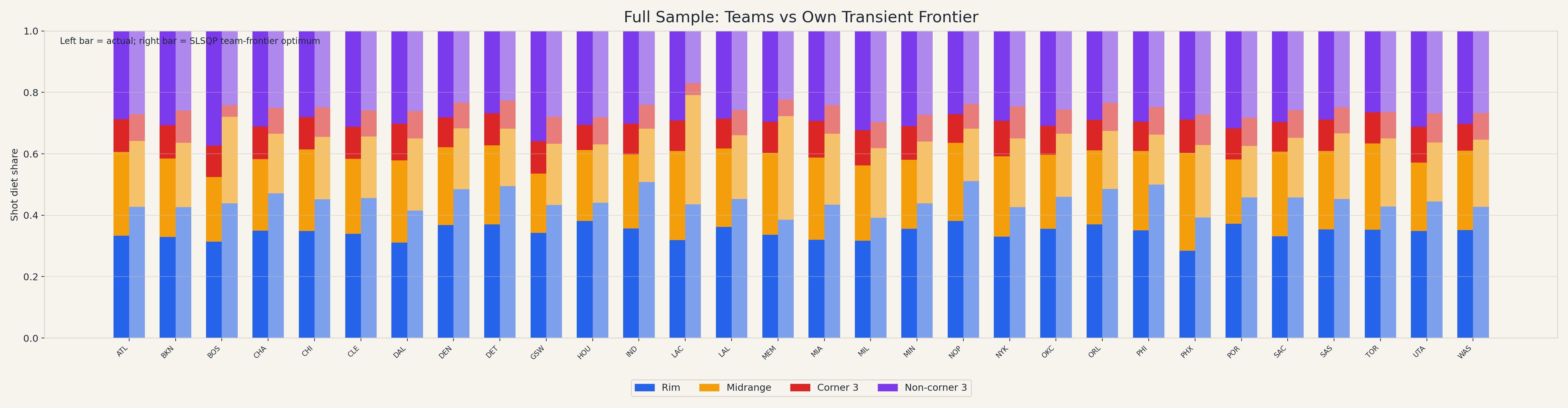}
\caption{Full-sample team shot diets compared with team-specific transient frontier portfolios.  This is also a full-sample optimization over all possessions, not the hybrid fixed-fastbreak recomposition.}
\label{fig:appendix-full-team-bars}
\end{figure}

Figures \ref{fig:appendix-full-league-bars} and \ref{fig:appendix-full-team-bars} are therefore different from the hybrid figures in the main text.  They do not keep fastbreaks fixed as an observed component; they optimize over the entire possession sample, including fastbreak possessions.

\section{Transition-Matrix Shrinkage Sensitivity}

The main team figures use empirical team transition matrices.  As a robustness check, we also compared those team matrices with an empirical-Bayes version that shrinks each team row toward the league transition matrix.  For team $a$, the smoothed transition probability is
\[
P^{(a,K)}_{ij}
=
\frac{n^{(a)}_{ij}+K P^{league}_{ij}}
{N^{(a)}_i+K},
\]
where $n^{(a)}_{ij}$ is the team's observed transition count from state $i$ to state $j$, $N^{(a)}_i=\sum_j n^{(a)}_{ij}$ is the team's row total, and $P^{league}_{ij}$ is the league transition probability.  The empirical matrix is the special case $K=0$:
\[
P^{(a,0)}_{ij}=\frac{n^{(a)}_{ij}}{N^{(a)}_i}.
\]

The motivation for the smoothed version is row-level stability.  Total team transition counts are large, but some specific rows are still uncommon after splitting by shot family, result, offensive-rebound source, and fastbreak status.  A positive $K$ acts like league pseudo-counts in those sparse rows.  We used $K=1000$ as the comparison value because it was the largest candidate in the validation grid and had the lowest validation mean-squared error in the no-fastbreak transition validation.  It is also small relative to each team's total five-season transition count, so it is a stability check rather than a replacement of team identity.

\begin{table}[H]
\centering
\caption{K=0 versus K=1000 team-frontier comparison.}
\label{tab:k-sensitivity}
\resizebox{\textwidth}{!}{%
\begin{tabular}{lrrrrrr}
\toprule
Sample & Teams & Same diets & Mean L1 diet dist. & Max L1 diet dist. & Mean $|\Delta V|$ & Max $|\Delta$ gap$|$ \\
\midrule
Hybrid recomposition & 30 & 0 & 0.0525 & 0.8575 & 0.0080 & 0.1342 \\
No fastbreak & 30 & 0 & 0.0587 & 0.9545 & 0.0080 & 0.1342 \\
\bottomrule
\end{tabular}
}
\end{table}

Both the hybrid and no-fastbreak comparisons are usually close, but each has at least one team-level result that is sensitive to smoothing.  The hybrid comparison keeps observed fastbreak shot selection fixed and changes only the no-fastbreak team frontier from $K=0$ to $K=1000$.  This is why the main text uses the empirical team matrices and keeps the shrinkage version as a sensitivity check rather than as the main team result.

\begin{figure}[H]
\centering
\includegraphics[width=\textwidth]{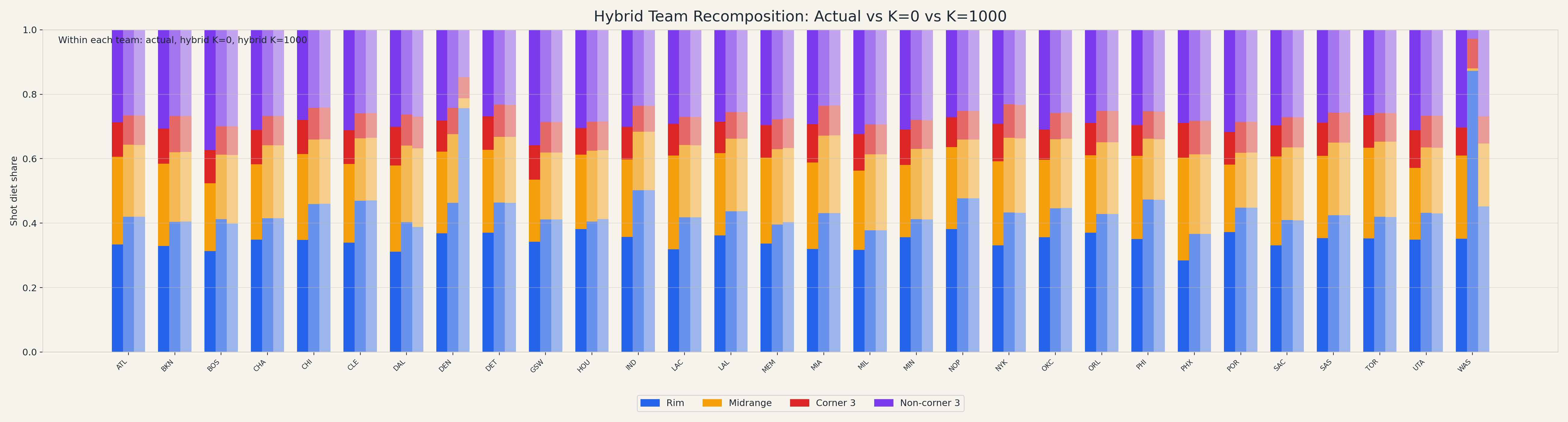}
\caption{Full actual team diets compared with hybrid recompositions using the no-fastbreak $K=0$ and $K=1000$ team-frontier diets.  Fastbreak shot selection is held fixed from observed removed fastbreak possessions.}
\label{fig:appendix-hybrid-k-three-bars}
\end{figure}

\begin{figure}[H]
\centering
\includegraphics[width=\textwidth]{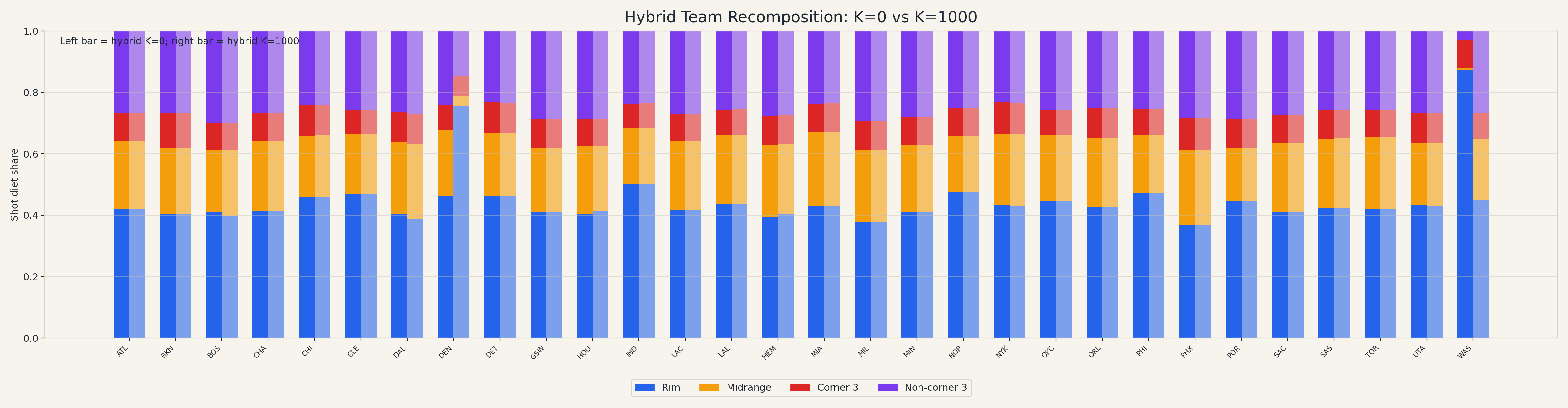}
\caption{Hybrid team-frontier diets compared directly under $K=0$ and $K=1000$.}
\label{fig:appendix-hybrid-k-two-bars}
\end{figure}

\begin{figure}[H]
\centering
\includegraphics[width=\textwidth]{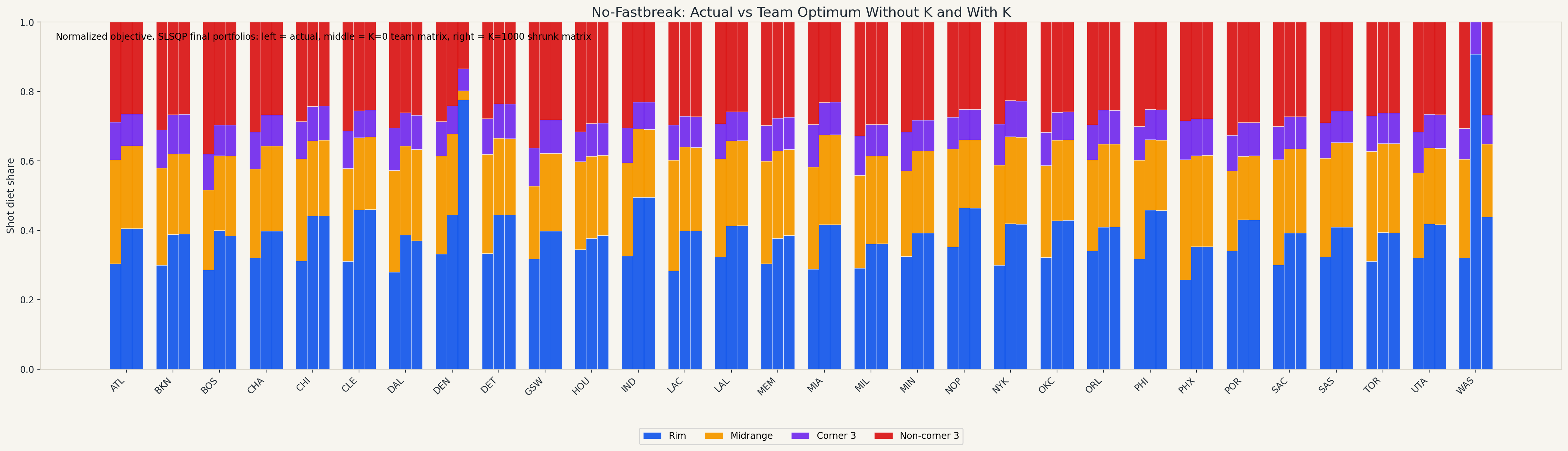}
\caption{No-fastbreak team actual diets compared with empirical team-frontier diets and $K=1000$ team-frontier diets.}
\label{fig:appendix-nf-k-three-bars}
\end{figure}

\begin{figure}[H]
\centering
\includegraphics[width=\textwidth]{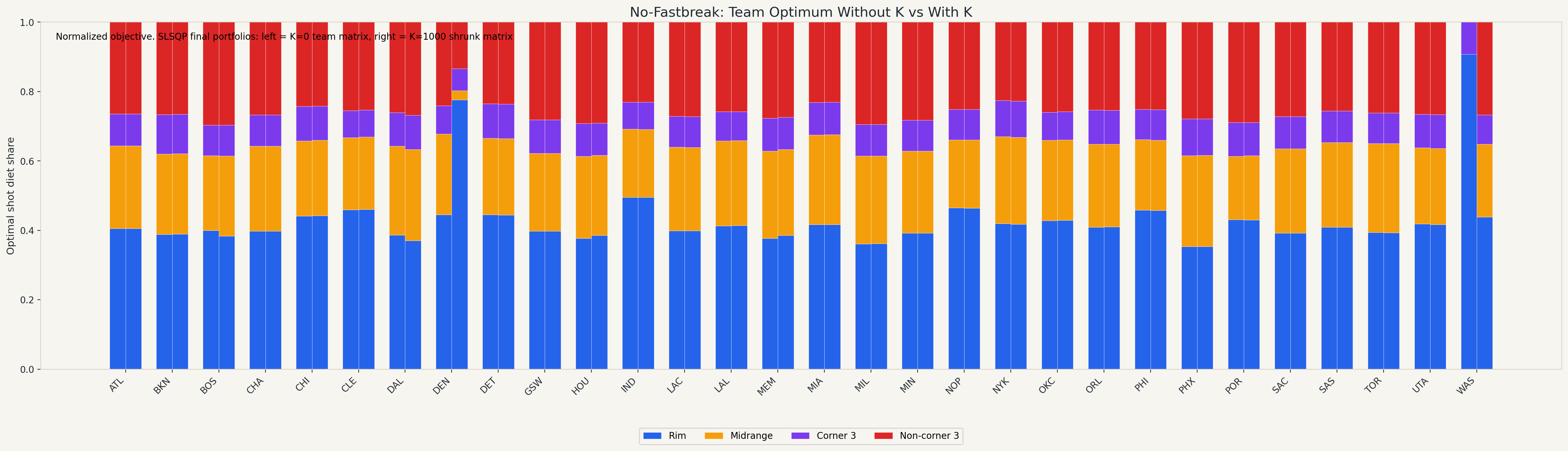}
\caption{No-fastbreak empirical team-frontier diets compared directly with $K=1000$ team-frontier diets.}
\label{fig:appendix-nf-k-two-bars}
\end{figure}

\begin{figure}[H]
\centering
\includegraphics[width=0.72\textwidth]{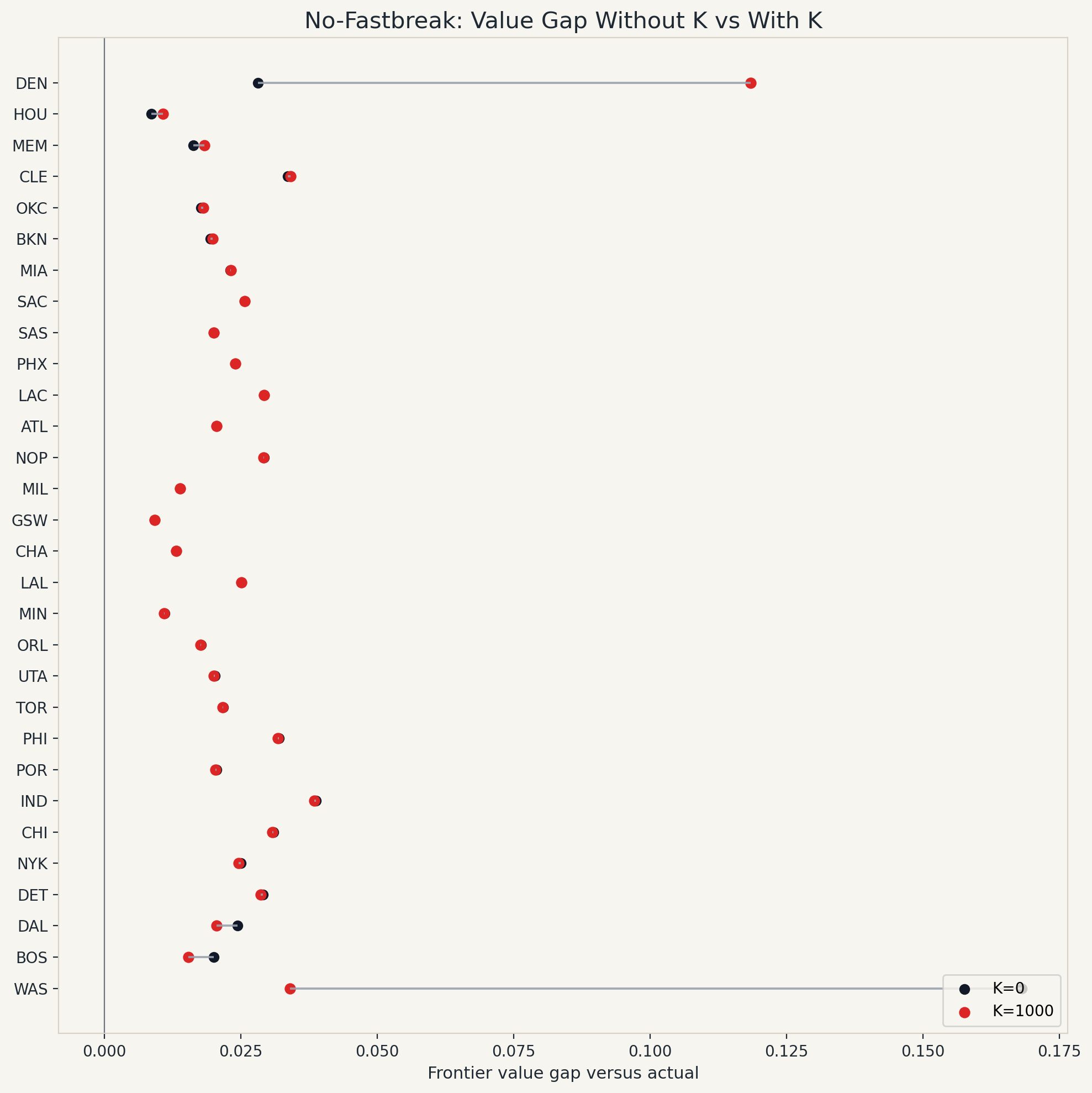}
\caption{No-fastbreak frontier value gaps under empirical and $K=1000$ team transition matrices.}
\label{fig:appendix-nf-k-gap}
\end{figure}

\end{document}